# Ultrafast carriers' separation imaging in WS$_2$-WSe$_2$ in plane heterojunction by transient reflectivity microscopy

Yangguang Zhong[1,2,†], Shuai Yue[1,†,*], Huawei Liu[2,†], Yuexing Xia[1], Anlian Pan[2*], Shula Chen[2*] and Xinfeng Liu[1*]

[1]CAS Key Laboratory of Standardization and Measurement for Nanotechnology, CAS Center for Excellence in Nanoscience, National Center for Nanoscience and Technology, Beijing 100190, China.

[2]Key Laboratory for Micro-Nano Physics and Technology of Hunan Province, State Key Laboratory of Chemo/Biosensing and Chemometriscs and College of Materials Science and Engineering, Hunan University, Changsha, Hunan 410082, China.

[†]These authors contributed equally: Yangguang Zhong, Shuai Yue, Huawei Liu.

[*]Email address: liuxf@nanoctr.cn, shuaiyue@nanoctr.cn, shuch@hnu.edu.cn, Anlian.pan@hnu.edu.cn

## Abstract

Carrier transport in nanodevices plays a crucial role in determining their functionality. In the post-Moore era, the behavior of carriers near surface or interface domains the function of the whole devices. However, the femtosecond dynamics and nanometer-scale movement of carriers pose challenges for imaging their behavior. Techniques with high spatial-temporal resolution become imperative for tracking their intricate dynamics. In this study, we employed transient reflectivity microscopy to directly visualize the charge separation in the atomic interface of WS$_2$-WSe$_2$ in-plane heterojunctions. The carriers' drifting behavior was carefully tracked, enabling the extraction of drift velocities of 30 nm/ps and 10.6 nm/ps for electrons and holes. Additionally, the width of the depletion layer was determined to be 300 nm based on the carriers' moving trajectory. This work provides essential parameters for the potential effective utilization of these covalent in-plane heterojunctions, and demonstrates the success of transient optical imaging in unraveling the electrical behavior of nano devices, paving the way for a new avenue of electro-optical analysis.

# Introduction

The field of electrical engineering critically relies on the precise control of carrier movement. Carrier transport imaging serving as a fundamental technique in understanding and optimizing the performance of semiconductor devices have attracted great attention. In the post Moore law period, the electrical devices continue to shrink, the behavior of devices predominantly governed by the interface, even single atom layers [1, 2, 3, 4, 5]. To investigate carrier dynamics at these extremely localized levels, advanced techniques with high spatial-temporal resolution are essential. Electron beams or scanning probes provide high spatial resolution, whereas pulsed laser beams excel in temporal resolution. In the quest for improved time, spatial, and energy resolutions, diverse testing systems have been devised, accomplishing both high spatial and temporal resolution. Examples include ultrafast electron microscopy and time-resolved photoemission electron microscopy, both offering nanometer-level resolution [6, 7, 8]. The latter has found extensive use in revealing carrier behaviors near interfaces. Nevertheless, electron microscopy exclusively detects electron-based phenomena, and its limitations, such as low resolution (10 meV), conductivity requirements of the sample, and the potential risk of damage, also restrict its application in precise energy transfer processes. Optical methods, such as transient absorption microscopy, benefit from high temporal and energy resolution, enabling nanoscale imaging of complex exciton/carrier or heat transport [9, 10, 11, 12]. This provides a powerful tool for in-depth exploration of material behavior at the microscopic level.

Among the numerous transistors, in-plane single layer heterojunction is a kind of smallest one which is composed of only several atoms. These structures have garnered significant attention for their potential applications in nano devices [13, 14, 15]. Unlike stacked transistors, the covalent bonding in in-plane single-layer heterojunctions is expected to result in much higher transport velocities. However, the detailed dynamics of carriers' transport at the interface in these structures are rarely studied [16, 17, 18].

To address this gap, we employed transient reflection microscopy (TRM) to investigate the basic carrier dynamics near the interface of a $WS_2$-$WSe_2$ heterojunction. This technique demonstrated a unique advantage in distinguishing transported carriers. Through tracking ultrafast charge separation and movement under the built-in electric field near the interface by selecting specific detection wavelengths the velocities of holes and electrons were extracted. Based on the movement behavior of carriers, the boundaries of depletion layer were also identified. These findings are essential for the design of devices utilizing in-plane single-layer heterojunctions, showcasing the potential of transient reflection microscopy in advancing our understanding of carrier dynamics in nanoscale materials.

## Results

**Basic characterization of lateral heterojunctions.** We investigated high-quality $WS_2$-$WSe_2$ lateral heterostructures, synthesized through chemical vapor deposition (CVD) on a $SiO_2$/Si substrate [3, 19], as the focal point of our study (**Figure 1A**). The central triangular region comprises $WS_2$, while the outer dark region corresponds to $WSe_2$. The sample exhibits a clean surface with a side length of approximately 50 μm, and the lateral width of $WSe_2$ is roughly 10 μm. Atomic force microscopy (AFM) image of the heterojunction confirms that the heterojunction is a single layer with a thickness of approximately 0.9 nm (**Figure S1**)[20]. Scanning transmission electron microscopy (STEM) provides exceptional resolution and can reveal the internal structure and morphology of materials, enabling researchers to identify individual atoms, defects, and other structural features. **Figure 1B** displays a high-resolution STEM (HRSTEM) image of the heterojunction region, interface site was indicated by the white dashed line. Due to the different scattering strength with electrons by W, S and Se atoms, we can observe the atomically sharp and covalently bonded zigzag interface between the $WSe_2$ and $WS_2$ in **Figure S2A** [21]. The up panel in **Figure S2B** shows the atomic structure of the heterostructures, the lower panel shows the magnification HRTEM image, these W, S, Se atoms marked as blue, yellow and green ball. **Figure 1C** demonstrate the Kelvin probe force microscopy (KPFM) image of the interface, the indicator white line across the junction shows a sharp voltage step, which indicates the strong build in electric field [21, 22, 23]. The absorption spectra of the junction region and single component are shown in **Figure 1D** The absorption peak of A exciton, B exciton and C exciton of $WS_2$ are 2.0 eV, 2.39 eV and 2.84 eV respectively. And the correspond absorption peak of $WSe_2$ are 1.67 eV, 2.08 eV and 2.45 eV. The excitonic A and B peaks observed in monolayers of $WS_2$ and $WSe_2$ arise from transitions occurring between the spin-orbit split valence band and the lowest conduction band situated at the K and K′ points within the Brillouin zone. The A exciton absorption peak of $WS_2$ and $WSe_2$ on interface remain a constant contrast with their single domain. However, the exciton absorption peaks of the spectrum of the heterojunction region are much weaker, indicating the strong dissociation of the excitons within depletion region [24]. The carriers were excited by photons with energy of 3.1 eV while probed by super continuum white light or monochromatic laser with energy matched with the exciton's absorption.

Photoluminescence (PL) line scan image across the interface is shown in **Figure 1E**, The PL line scan across the lateral interface showed in **Figure 1E**, which indicate that the position of the PL peak for the two domains remained constant within each region, alloy near the junction could be excluded [18]. The emission energy of the A exciton in $WS_2$ and $WSe_2$ domains were determined to be 1.96 eV and 1.64 eV eV in **Figure S3**, while on the interface, the peak of $WS_2$ and $WSe_2$ were determined to be 1.95 eV (-8 meV red shift) and 1.65 eV (14 meV blue shift), corresponds to recombination of their respective A excitons. And the blue and red shift of the peak which may cause by charge transfer between two domains. Previous work demonstrates a type-II band alignment [3, 25], which was shown in the **Figure 2F**, the band offset in the conduction band and valence band were estimated to be 0.37 eV and 0.65 eV [22]. The dash line indicates the Femi level of the two materials [25, 26].

Charge transfer dynamics of the interface. The interface of the lateral heterojunction is formed by covalent bonding of two different materials, each with distinct Fermi levels, creating a built-in electric potential [17, 27]. Our KPFM tests on the heterojunction yielded a built-in potential of 78 mV, a value consistent with literature reports [22, 28]. In **Figure 2A**, a schematic diagram illustrates carrier excitation and separation; when excited by photons, carriers within the depletion region are propelled by the built-in electric field. Positive carriers move towards the p-type region, while negative carriers migrate towards the n-type region[17]. Charge separation dynamics were experimentally demonstrated by comparing transient reflection signals at the interface and within single components. The presence of the built-in potential has a significant impact on photogenerated excitons/carriers at the interface, and unstable charge transfer (CT) excitons at the interface have been reported [18]. In the previous work, the binding energy of CT excitons is only 30 meV in the lateral heterojunction system encapsulated in h-BN [29]. The binding energy of CT is related to the interface width (L), dielectric environment ($\varepsilon$), and band gap difference ($\Delta E_v$ or $\Delta E_c$). A longer interface width, larger dielectric environment, and larger band gap difference will result in a smaller binding energy. Compared to corresponding parameters in the literature ($\varepsilon_{h-BN}$: ~4.5, $\Delta E_v$: ~215 meV), our system has a larger dielectric constant and band gap difference ($\varepsilon_{Al2O3}$: ~9, $\Delta E_v$: ~650 meV),[22] hence a smaller binding energy (less than the built-in potential). Therefore, photogenerated carriers at the interface will rapidly dissociate under the influence of the built-in potential. However, the detailed carriers' dynamics haven't been fully studied.

As optical measurements can detect excitons within single component and carriers transfer process across the interface [30, 31]. We first compared the carrier dynamics between them using micro-area transient reflection spectra, as shown in **Figures 2B** and **2C**. We conducted transient reflection spectra (TRS). In **Figure 2B**, the TRS signal contrast of the heterojunction (upper panel) and single WS$_2$ region (lower panel) is presented with a delay time of 0 - 100 ps at room temperature. The pump light is at 400 nm, and white light is used as the probe light with a wavelength of 500 - 775 nm. An exciton bleaching signals of WS$_2$ and WSe$_2$ are observed at 1.95 eV and 1.65 eV, respectively, accompanied with light-induced absorption signals. Additionally, positions of 2.38 eV and 2.09 eV correspond to the B-exciton bleaching signals of $WS_2$ and $WSe_2$, respectively. Finally, in **Figure 2C**, the decay kinetics of the A exciton signal in pure $WS_2$ and the $WS_2$ region at the interface are shown, with the lifetime of the $WS_2$ A excitons in the heterojunction region exhibiting a significant increase, attributed to the charge transfer at the interface [24, 32, 33, 34]. Additional TAS data are presented in **Figure S4**. The exciton dynamics of the two materials were extracted separately, revealing a noteworthy increase contributed by the transferred carriers. Micro-region transient absorption spectroscopy data unveil a distinct carrier transfer phenomenon at the interface, with electron and hole transfer efficiencies measured at approximately 71.8% and 46.7%, respectively [35].

**Transient absorption microscopy of the interface.** The built-in electric field present in the lateral heterojunction will quickly dissociate the photo-generated excitons. However, this crucial process hasn't been imaged in experiment yet. Here, we successfully captured this charge separation process using transient reflection microscopy [9, 10]. **Figure 3A** illustrates the diagram of transient reflective imaging of charge separation at the heterojunction interface. Optical-induced carriers were excited by a focused pump beam with an energy of 3.1 eV and detected by probe beams with energies of 2.01 eV (615 nm) and 1.71 eV (724 nm). The probe beams were pre-focused to achieve wide-field detection on the sample. Time delays between the pump and probe beams, ranging from -2 ps to 30 ps with a resolution of 0.2 ps, were controlled by a motorized stage. Carrier-induced reflection changes at controllable time delays were detected by a CMOS camera. To enhance the signal-to-noise ratio, a shot-to-shot measurement strategy was implemented. The pump beam was chopped at a frequency half of the repetition rate of the laser source, and the reflected probe beams with and without pump excitation were recorded alternately by the CMOS for more than 100,000 times. A small fraction of the chopped pump beam served as a reference signal to identify the sign of the signal. The schematic diagram of the TAM optical system is shown in **Figure S5**[10].

The drifting movement of carriers is illustrated in **Figures 3B** and **3C**. Under the influence of the built-in electric field, dissociated electrons and holes drift towards the negative and positive regions, respectively, until they reach the depletion layer boundary[17]. By using probe beams sensitive only to the P or N region, we directly captured this charge separation at the interface. The 2.01 eV laser beam was chosen for the N region, and the 1.71 eV laser beam was selected for the P region. **Figure 3B** displays the carrier movement with N-type sensitivity imaging at the heterojunction within 5 ps. A clear carriers' movement could be captured from the near interface area to the outer side. A red dotted line represents the position of the heterojunction interface. Yellow circles mark the carrier's distribution at different delay times and a white dotted arrow connects the peak positions at different moments, revealing an obvious linear movement trajectory. **Figure 3C** presents the imaging of P-type sensitive carrier movement (in red), showing a continuous but slower movement which indicative of a much heavier effective mass of holes[8]. A comparison of the dynamics of the pure components of $WS_2$ and $WSe_2$ are presented in **Figure S6** and **S7**. In contrast to the movement of the signal peak at the interface, no obvious movement could be detected in the pure components.

## Discussion

To assess the precise exciton separation and carrier transport at the interface, we utilized a drift-diffusion model to simulate the spatiotemporal evolution of the excited carrier population across the junction. This model considers carriers evolving within a potential landscape $U(x)$ and decaying via a single molecular recombination rate of $k$, with the carrier population, $n\,(r,\,t)$, governed by the equation [16, 18]:

$$\frac{\partial n}{\partial t} = D\nabla^2 n(t) + \frac{D}{k_B T}\nabla(n(t)\nabla U) - kn \;,$$

where $D$ is the diffusion coefficient in the monolayer. We model $U$ as a smooth step function,

$$U(x) = \frac{U_0}{2}\left\{1 - erf(\frac{x-w}{\sqrt{2}\sigma})\right\},$$

Where *erf* is the error function, such that $\nabla U$ is a Gaussian of variance $\sigma$ [18].

Slits of TRM of carriers at the interface in long periods are presented in **Figure 4** (with detailed TRM data shown in **Figure S8** and **S9**). **Figure 4A** illustrates profiles detected at 2.01 eV, with the red dashed line denoting the interface. Gaussian fitting was employed to determine the central peak position of the profile (**Figure 4B**). Time-delay-dependent peak positions were extracted and are depicted in **Figure 4C**. Within the initial 5 ps, the peak position shifted from 0 to 150 nm, remaining relatively stable thereafter. Linear fitting yielded a slope of 30 nm/ps, corresponding to a speed of electrons driven from the higher Fermi level of $WSe_2$ to $WS_2$ under the built-in electric field [22]. As the electrons reached the depletion layer boundary, their speed significantly decreased, and the width of the half-depletion region was determined to be 150 nm. In contrast, isolated $WS_2$ and $WSe_2$ exhibited a relatively unchanged TRM slit, as depicted in **Figure S10** and **S11**. Similarly, when using 1.71 eV as the detection energy at the interface, the drift motion of holes was observed (**Figure 4D-E**). Within 15 ps, the position linearly moved to a maximum of 150 nm and then became static. Linear fitting yielded a drift velocity of 10.6 nm/ps, corresponding to the hole drift velocity. Accordingly, the width of the depletion layer in the $WSe_2$ region was determined to be 150 nm, resulting in a total depletion layer width of 300 nm. In summary, we observed the drift motion of electrons and holes in the depletion layer under the built-in electric field, determining carrier drift velocities and the width of the interface depletion layer.

## Conclusions

In conclusion, our study presents a detailed characterization of a monolayer $WS_2$-$WSe_2$ lateral heterojunction, offering a comprehensive investigation into the carrier properties and drifting movements within lateral heterojunctions of $WSe_2$ and $WS_2$. This research highlights the unique properties of these heterojunctions and underscores their potential applications. Our work contributes to the ongoing efforts in advancing the field of 2D materials and their heterojunctions, providing important insights into the design and optimization of high-performance electronic devices based on lateral heterojunctions. Additionally, we establish that TRM is a powerful optical technique for detecting and studying carriers' transport dynamics, further enhancing the toolkit for investigating semiconductor behavior in nanoscale materials.

**Materials and Methods**

*Lateral heterostructure synthesis.* The $WS_2$/$WSe_2$ heterojunction nanosheets were synthesized by a source switching vapor growth strategy. An alumina boat with $WS_2$ powder (99.999%, Alfa Aesar) was located at central of the furnace and another boat with WSe2 powder (99.999%, Alfa Aesar) was put at the upstream out of the heating zone. The two boats are separated by a quartz rod, and a step motor was used to control the position of these boats in the tube through the magnetic force system. Several $SiO_2$/Si substrates were placed downstream of the gas flow, 10 cm far away from the furnace center. Before heating, the system was flushed by high pure nitrogen for 10 mins at a rate of 150 sccm. Then the center temperature of the furnace was ramped to 980 °C with a rate of 14 °C/min, and the $WS_2$ nanosheets were achieved after 2 mins growth. For the epitaxy growth of $WSe_2$, the temperature of the furnace was descended to 950 °C with a rate of 10 °C/min and then pushing the alumina boat with the $WSe_2$ powder into the central of the furnace quickly to replace the $WS_2$ boat. After 2 mins growth the system cooling down naturally, and the $WS_2$/$WSe_2$ heterojunction nanosheets can be achieved finally.

*Scanning Transmission Electron Microscopy.* Monolayer lateral heterostructures was transfer to the Cu grids by wet transfer method. The lateral heterostructures was studied by TEM and HRTEM using a TECNAI G2F30 S-TWIN transmission electron microscope operating at 300 kV.

*Photoluminescence and Raman measurement.* PL and Raman spectra were acquired using a confocal micro-Raman module (Institute of Semiconductors, CAS) coupled to a Horiba iHR550 imaging spectrometer and a 100X objective lens (Olympus). Both PL and Raman spectra measurements all used a 532 nm laser as the exciting source. PL and Raman mapping of lateral heterostructures was obtained using a the electrically controlled stage by the step size 0.5 μm both *x* and *y* direction.

*Steady state reflection absorption spectrum.* For the steady state reflection absorption spectra measurements of the monolayer pn junction, the white light was focused on the sample by a 100× Olympus objective lens with NA~0.9, and the target area of the sample was selected by the pinhole. The reflection spectrum of the sample, marked as $I_{sam}$, was measured. Then the reflection spectrum of the substrate, marked as $I_{sub}$, was measured. The absorption spectrum of the sample was obtained: $(I_{sub} - I_{sam})/ I_{sub}$.

*Pump-probe reflection spectroscopy and Transient reflection microscopy.* A commercial transient absorption spectrometer module (TA100, Time-Tech Spectra) was used. Use 800 nm pulsed light as the fundamental wave pulse (Chameleon, 80 MHz, 800 nm, 3.5W, 680-1080 nm). After passing through the beam splitter, the part passes through the hollow angle mirror fixed on the delay line, and then uses BBO frequency multiplication to generate 400 nm as pump source; the other part of the fundamental frequency light passes through the white light crystal to produce super continuous white light with a wavelength range of 490 nm-775 nm as probe light. Microscopic TA measurements of monolayer p-n junction was performed using a home-made microscopy system integrated into the TA100. The time trace of the oscillation component at the detection wavelength is obtained by subtracting the overall dynamics from the transient spectrum. The transient reflectivity microscopy is the same as the system in the previous work of our group. (9) For the 400 nm pump, 615/724 nm probe experiments, a β-Barium borate (BBO) crystal was used to convert partial of 800 nm laser to 400 nm, while OPA was used to generate detection pulse. Due to the influence of the $SiO_2$/Si substrate on the detection light, the lateral heterojunction was transferred to the C-plane transparent sapphire substrate, and all TAS and TAM test were performed based on samples on the sapphire substrate.

**Associated content**

*Supplementary Information

The Supplementary Information is available.

Figure S1-S11 and calculation as described in the text.

**Author information**

Corresponding Authors

*Email address: liuxf@nanoctr.cn, shuaiyue@nanoctr.cn, shuch@hnu.edu.cn, Anlian.pan@hnu.edu.cn

**Author contributions:** X.L., S.Y., S.C. and A.P. lead the project. S.Y., Y.Z. and Y.X. construct the optical set up. H.L. produced the samples and provided the selected area electron diffraction data of the sample. Y.Z. and S.Y. perform the optical spectroscopy and data analysis. Y.Z., S.Y. and X.L. prepared the manuscript. All the authors discussed the results and revised the manuscript.

**Acknowledgments**

**Funding:** The authors acknowledge the National Key Research and Development Program (2023YFA1507002,2022YFA1603701), the National Natural Science Foundation of China (22173025, 22073022, 62175061, 52172140, 52221001), National Science Foundation for Distinguished Young Scholars of China (22325301). Youth Innovation Promotion Association CAS. The Natural Science Foundation of Hunan Province (2022JJ30167). The National Key R&D Program of China (No. 2022YFA1204300); The National Natural Science Foundation of China (Nos. 52221001, 62090035).

**Competing interests:** Authors declare that they have no competing interests.

**References**
1. Geim AK, *et al.* Van der Waals heterostructures. *Nature* **499**(7459)**:** 419-425.(2013).
2. Novoselov KS, *et al.* 2D materials and van der Waals heterostructures. *Science* **353**(6298)**:** aac9439.(2016).
3. Zhang ZW, *et al.* Robust epitaxial growth of two-dimensional heterostructures, multiheterostructures, and superlattices. *Science* **357**(6353)**:** 788-792.(2017).
4. Akinwande D, *et al.* Graphene and two-dimensional materials for silicon technology. *Nature* **573**(7775)**:** 507-518.(2019).
5. Sun X, *et al.* Enhanced interactions of interlayer excitons in free-standing heterobilayers. *Nature* **610**(7932)**:** 478-484.(2022).
6. Yang DS, *et al.* Scanning ultrafast electron microscopy. *PNAS* **107**(34)**:** 14993-14998.(2010).
7. Cho JW, *et al.* Visualization of carrier dynamics in p(n)-type GaAs by scanning ultrafast electron microscopy. *PNAS* **111**(6)**:** 2094-2099.(2014).
8. Najafi E, *et al.* Four-dimensional imaging of carrier interface dynamics in p-n junctions. *Science* **347**(6218)**:** 164-167.(2015).
9. Delor M, *et al.* Imaging material functionality through three-dimensional nanoscale tracking of energy flow. *Nat. Mater.* **19**(1)**:** 56-62.(2020).
10. Yue S, *et al.* High ambipolar mobility in cubic boron arsenide revealed by transient reflectivity microscopy. *Science* **377**(6604)**:** 433-436.(2022).
11. Del Aguila AG, *et al.* Ultrafast exciton fluid flow in an atomically thin $MoS_2$ semiconductor. *Nat. Nanotechnol.* **18**(9)**:** 1012-1019.(2023).
12. Zhang Q, *et al.* Interface nano-optics with van der Waals polaritons. *Nature* **597**(7875)**:** 187-195.(2021).
13. Duan XD, *et al.* Lateral epitaxial growth of two-dimensional layered semiconductor heterojunctions. *Nat. Nanotechnol.* **9**(12)**:** 1024-1030.(2014).
14. Gong YJ, *et al.* Vertical and in-plane heterostructures from $WS_2$/$MoS_2$ monolayers. *Nat. Mater.* **13**(12)**:** 1135-1142.(2014).
15. Huang CM, *et al.* Lateral heterojunctions within monolayer $MoSe_2$-$WSe_2$ semiconductors. *Nat. Mater.* **13**(12)**:** 1096-1101.(2014).
16. Gauriot N, *et al.* Direct Imaging of Carrier Funneling in a Dielectric Engineered 2D Semiconductor. *ACS Nano* **18**(1)**:** 264-271.(2023).
17. Massicotte M, *et al.* Dissociation of two-dimensional excitons in monolayer $WSe_2$. *Nat. Commun.* **9**(1)**:** 1633.(2018).
18. Yuan L, *et al.* Strong Dipolar Repulsion of One-Dimensional Interfacial Excitons in Monolayer Lateral Heterojunctions. *ACS Nano* **17**(16)**:** 15379-15387.(2023).
19. Li MY, *et al.* Epitaxial growth of a monolayer $WSe_2$-$MoS_2$ lateral p-n junction with an atomically sharp interface. *Science* **349**(6247)**:** 524-528.(2015).
20. Shi J, *et al.* Twisted-Angle-Dependent Optical Behaviors of Intralayer Excitons and Trions in $WS_2$/$WSe_2$ Heterostructure. *ACS Photonics* **6**(12)**:** 3082-3091.(2019).
21. Sahoo PK, *et al.* One-pot growth of two-dimensional lateral heterostructures via sequential edge-epitaxy. *Nature* **553**(7686)**:** 63-67.(2018).
22. Zheng BY, *et al.* Band Alignment Engineering in Two-Dimensional Lateral Heterostructures. *J. Am. Chem. Soc.* **140**(36)**:** 11193-11197.(2018).
23. Wan X, *et al.* Gate-Tunable Junctions within Monolayer $MoS_2$–$WS_2$ Lateral Heterostructures. *ACS Appl. Nano Mater.* **5**(10)**:** 15775-15784.(2022).
24. Wang K, *et al.* Interlayer Coupling in Twisted $WSe_2$/$WS_2$ Bilayer Heterostructures Revealed by Optical Spectroscopy. *ACS Nano* **10**(7)**:** 6612-6622.(2016).
25. Kang J, *et al.* Band offsets and heterostructures of two-dimensional semiconductors. *Appl. Phys. Lett.* **102**(1).(2013).

26. Cao Z, *et al.* Impact of Interfacial Defects on the Properties of Monolayer Transition Metal Dichalcogenide Lateral Heterojunctions. *Journal of Physical Chemistry Letters* **8**(7)**:** 1664-1669.(2017).
27. Zhang CD, *et al.* Strain distributions and their influence on electronic structures of $WSe_2$-$MoS_2$ laterally strained heterojunctions. *Nat. Nanotechnol.* **13**(2)**:** 152-158.(2018).
28. Zheng BY, *et al.* Two-Dimensional Lateral Multiheterostructures Possessing Tunable Band Alignments. *Chem. Mater.* **35**(17)**:** 6745-6753.(2023).
29. Rosati R, *et al.* Interface engineering of charge-transfer excitons in 2D lateral heterostructures. *Nat. Commun.* **14**(1)**:** 2438.(2023).
30. Bellus MZ, *et al.* Photocarrier Transfer across Monolayer $MoS_2$-$MoSe_2$ Lateral Heterojunctions. *ACS Nano* **12**(7)**:** 7086-7092.(2018).
31. Shimasaki M, *et al.* Directional Exciton-Energy Transport in a Lateral Heteromonolayer of $WSe_2$-$MoSe_2$. *ACS Nano* **16**(5)**:** 8205-8212.(2022).
32. Hong X, *et al.* Ultrafast charge transfer in atomically thin $MoS_2$/$WS_2$ heterostructures. *Nat Nanotechnol* **9**(9)**:** 682-686.(2014).
33. Rivera P, *et al.* Observation of long-lived interlayer excitons in monolayer $MoSe_2$-$WSe_2$ heterostructures. *Nat. Commun.* **6:** 6242.(2015).
34. Chen HL, *et al.* Ultrafast formation of interlayer hot excitons in atomically thin $MoS_2$/$WS_2$ heterostructures. *Nat. Commun.* **7:** 12512.(2016).
35. Peng B, *et al.* Ultrafast charge transfer in $MoS_2$/$WSe_2$ p–n Heterojunction. *2D Materials* **3**(2)**:** 025020.(2016).

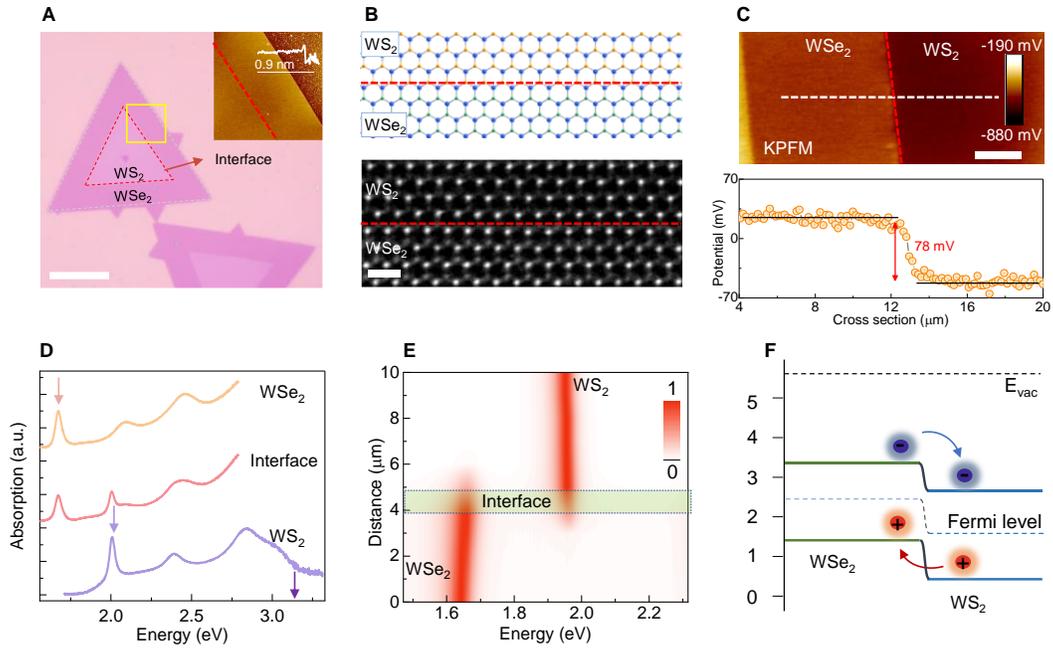

**Figure 1 | Optical characterization of lateral heterojunctions.** (**A**), Optical image of $WS_2$-$WSe_2$ monolayer lateral junction, the light-colored triangle marked by red dash line in the center is single-layer $WS_2$, and the dark-colored outside is the outline of single-layer $WSe_2$, the scale bar is 20 μm; the AFM image showed in the figure, dash red line indicates the interface and the height is ~0.9 nm. (**B**), Schematic of lateral heterostructure interface (upper panel) and atomic-resolution STEM image of the $WS_2$-$WSe_2$ lateral heterostructure (lower panel). The red dashed line indicates the atomic-scale zigzag edge interface, the scalebar is 0.5 μm. (**C**), KPFM image of the interface, the indicator white line across the junction shows a sharp voltage step, the unit is mV, result a built-in electric field ~78 mV. (**D**), absorption spectra of $WS_2$, $WSe_2$ and heterojunction regions, marked as $WS_2$, $WSe_2$ and HJs in the figure, the indicator present the probe and pump energy. (**E**), PL line scan image across the interface displayed in A. The blue line marked the peak energy of $WSe_2$ and $WS_2$, and the black dashed line indicates the interface. (**F**), Schematic of the theoretically predicted band alignment of a $WS_2$/$WSe_2$ heterostructure, which forms a type II heterojunction. Optical excitation of the heterostructure will lead to layer-separated electron (e-) and hole (h+) carriers, and the dash line indicate the fermi level.

Page **14** of **18**

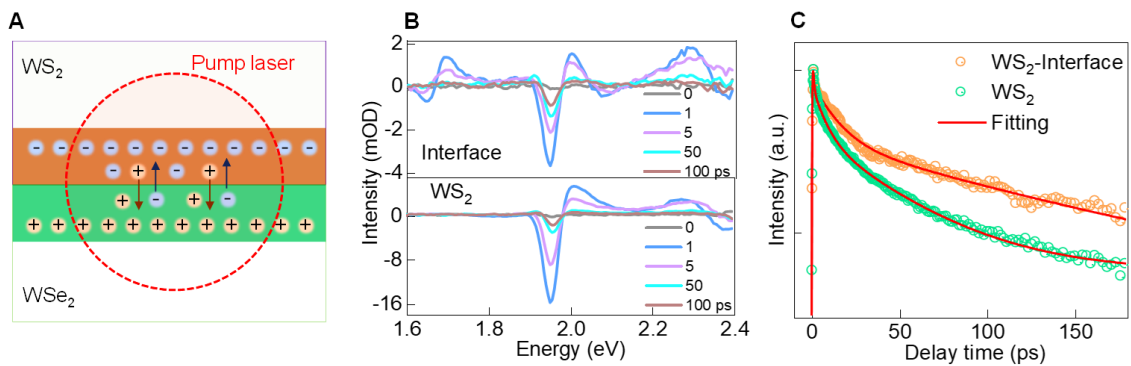

**Figure 2 | Charge transfer dynamics of the interface.** (**A**), Schematic diagram of carrier separation and diffusion at the light-excited heterojunction interface under the built-in electric field. (**B**), Micro TRS of the interface and pure $WS_2$ at different time delays from 0 to 100 ps. (**C**), the decay kinetics of the A exciton of $WS_2$ compared at pure domains and heterojunctions interface.

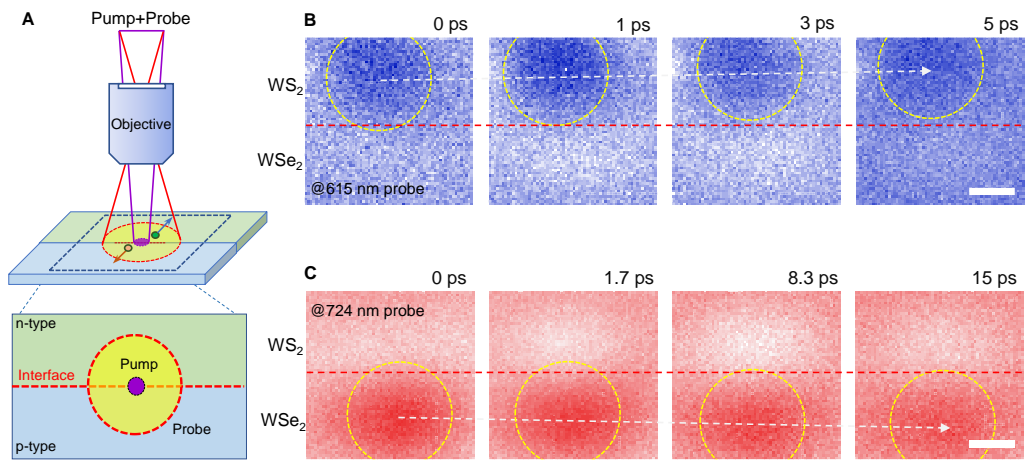

**Figure 3 | Transient absorption microscopy of the interface. (A)**, Schematic of the pump−probe microscopy experiment. **(B)**, Time evolution of the transient reflection signal after excitation on the interface probe by 615 nm. **(C)**, Time evolution of the transient reflection signal after excitation on the interface probe by 724 nm, all the scale bar are 0.5 μm. The red dash line indicates the site of the interface.

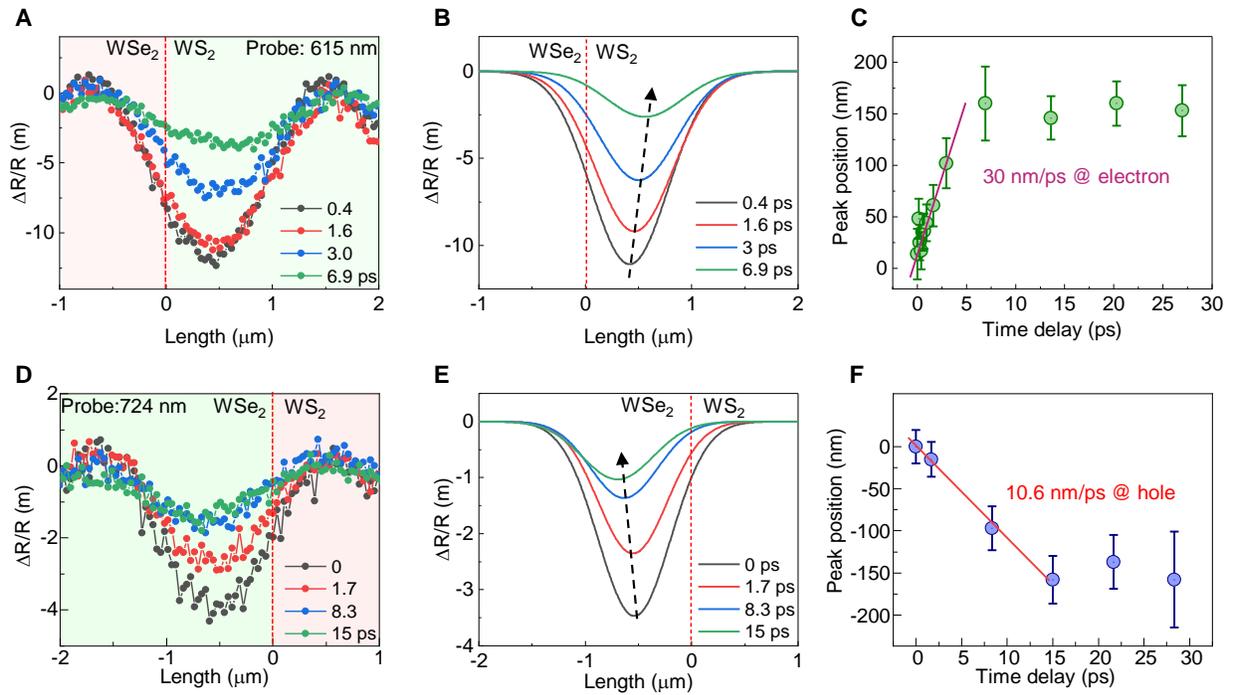

**Figure 4 | Slits of TRM at heterojunction region.** **(A)** Slits of TRM with the probe beam wavelength of 615 nm. **(B)** Gaussian fitting of figure **A**. **(C)** Peak position trajectory at different time delay **(D-E)** Slits of TRM with the probe beam wavelength of 724 nm and the related peak position tracking.

**TOC graphic**

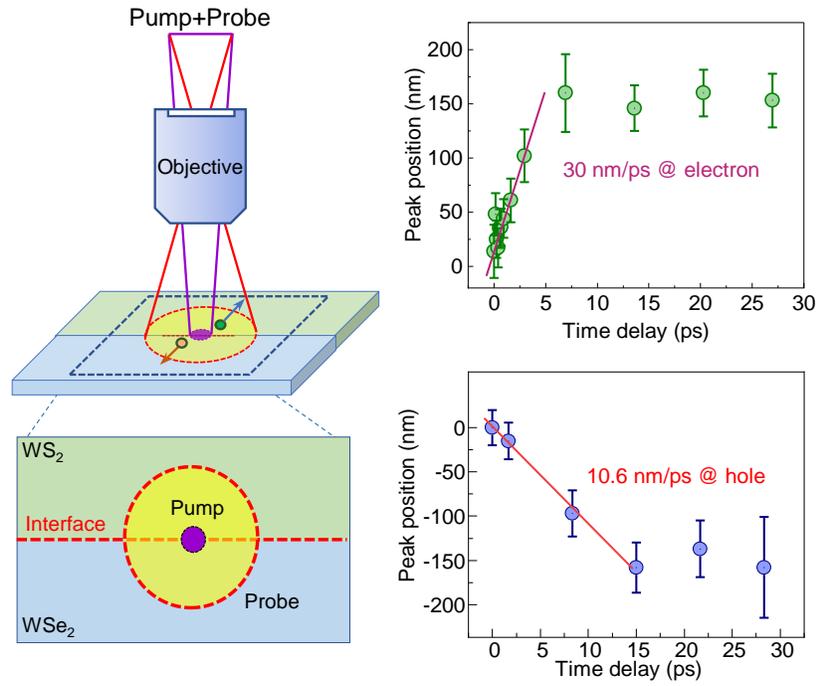